\documentclass[%
 reprint,
superscriptaddress,
 amsmath,
 amssymb,
aps,
pra,
showkeys
]{revtex4-2}
\usepackage{lineno,hyperref}
\usepackage[squaren]{SIunits}
\usepackage{color}
\usepackage{amsmath}
\usepackage{caption,subcaption}
\usepackage{setspace}
\usepackage{multirow}
\usepackage{threeparttable}
\usepackage{booktabs}
\usepackage{graphicx}
\usepackage{dcolumn}
\usepackage{bm}
\usepackage{float}
\usepackage{tabularx}

\definecolor{americanrose}{rgb}{1.0, 0.01, 0.24}
\definecolor{coralpink}{rgb}{0.97, 0.51, 0.47}
\definecolor{ao(english)}{rgb}{0.0, 0.5, 0.0}
\definecolor{darkpastelgreen}{rgb}{0.01, 0.75, 0.24}
\definecolor{navy}{rgb}{0.01, 0.0, 0.52}
\definecolor{cyan(process)}{rgb}{0.0, 0.72, 0.92}
\definecolor{brown}{rgb}{0.8, 0.5, 0}

\newcommand\si[1]{{\color{navy} {#1}}}

\begin{document}

\preprint{APS/123-QED}
\title{Predicting Accurate X-ray Absorption Spectra for CN$^+$, CN, and CN$^-$: Insights from Multiconfigurational and Density Functional Simulations}

\author{Jinyu Li}
 \affiliation{MIIT Key Laboratory of Semiconductor Microstructure and Quantum Sensing, Department of Applied Physics, School of Physics, Nanjing University of Science and Technology, 210094 Nanjing, China}

 \author{Sheng-Yu Wang}
   \email{wangshengyu@njust.edu.cn}
 \affiliation{MIIT Key Laboratory of Semiconductor Microstructure and Quantum Sensing, Department of Applied Physics, School of Physics, Nanjing University of Science and Technology, 210094 Nanjing, China}

\author{Lu Zhang}
 \affiliation{MIIT Key Laboratory of Semiconductor Microstructure and Quantum Sensing, Department of Applied Physics, School of Physics, Nanjing University of Science and Technology, 210094 Nanjing, China} 

 \author{Guoyan Ge}
 \affiliation{MIIT Key Laboratory of Semiconductor Microstructure and Quantum Sensing, Department of Applied Physics, School of Physics, Nanjing University of Science and Technology, 210094 Nanjing, China}

\author{Minrui Wei}
 \affiliation{MIIT Key Laboratory of Semiconductor Microstructure and Quantum Sensing, Department of Applied Physics, School of Physics, Nanjing University of Science and Technology, 210094 Nanjing, China} 

\author{Junxiang Zuo}%
 \email{jxzuo@njust.edu.cn}
 \affiliation{MIIT Key Laboratory of Semiconductor Microstructure and Quantum Sensing, Department of Applied Physics, School of Physics, Nanjing University of Science and Technology, 210094 Nanjing, China}
 
\author{Weijie Hua}%
 \email{wjhua@njust.edu.cn}
 \affiliation{MIIT Key Laboratory of Semiconductor Microstructure and Quantum Sensing, Department of Applied Physics, School of Physics, Nanjing University of Science and Technology, 210094 Nanjing, China}

\date{\today}
\begin{abstract}   

High-resolution X-ray spectroscopy is an essential tool in X-ray astronomy, enabling detailed studies of celestial objects and their physical and chemical properties. However, comprehensive mapping of high-resolution X-ray spectra for even simple interstellar and circumstellar molecules is still lacking. In this study, we conducted systematic quantum chemical simulations to predict the C1s X-ray absorption spectra of CN$^+$, CN, and CN$^-$. Our findings provide valuable references for both X-ray astronomy and laboratory studies. We assigned the first electronic peak of CN$^+$ and CN to C1s $\rightarrow \sigma^*$ transitions, while the peak for CN$^-$ corresponds to a C1s $\rightarrow \pi^*$ transition. We explained that the two-fold degeneracy ($\pi^*_{xz}$ and $\pi^*_{yz}$) of the C1s$\rightarrow\pi^*$ transitions results in CN$^-$ exhibiting a significantly stronger first absorption compared to the other two systems. We further calculated the vibronic fine structures for these transitions using the quantum wavepacket method based on multiconfigurational-level, anharmonic potential energy curves, revealing distinct energy positions for the 0-0 absorptions at 280.7 eV, 279.6 eV, and 285.8 eV. Each vibronic profile features a prominent 0-0 peak, showing overall similarity but differing intensity ratios of the 0-0 and 0-1 peaks. Notably, introducing a C1s core hole leads to shortened C-N bond lengths and increased vibrational frequencies across all species. These findings enhance our understanding of the electronic structures and X-ray spectra of carbon-nitrogen species, emphasizing the influence of charge state on X-ray absorptions.


  \end{abstract}


\maketitle

\section{Introduction} \label{sec:intro}

The study of high-resolution X-ray spectra is essential for understanding the electronic  and chemical structures of molecules and ions, which are critical in the fields of molecular physics, chemistry, and astronomy.\cite{carravetta_x-ray_2022, ueda_high-resolution_2003, hergenhahn_vibrational_2004, svensson_soft_2005, bambi2023high, ezoe_high-resolution_2021} Among various X-ray techniques, vibrationally-resolved high-resolution X-ray absorption spectroscopy (XAS) has garnered significant interest from both experimental and theoretical researchers, covering a range of diatomic to polyatomic molecules and molecular ions.\cite{carravetta_x-ray_2022, mosnier_inner-shell_2016, lindblad_x-ray_2020, couto_breaking_2021, carniato_vibrationally_2020, zhang_franck-condon_2024}

Diatomic systems, in particular, serve as excellent models for benchmarking new theories (assessing the accuracy of methods, sensitivity to approximations or theoretical parameters, etc.) and investigating the influence of physical factors (charge state, bond length, external perturbations, etc.).  Within this context, CN$^+$, CN, and CN$^-$ are intriguing examples for investigating the influence of charge state on X-ray spectra. These species are interconnected through the ionization of the first and second valence electrons from CN$^-$:
\begin{eqnarray}
&&\text{CN}^- \rightarrow \text{CN} + e^- , \label{eq:i1} \\ 
&&\text{CN} \rightarrow \text{CN}^+ + e^-.\label{eq:i2}
\end{eqnarray}
The influence of charge state on X-ray photon spectra (XPS), or electron spectroscopy for chemical analysis (ESCA), has been extensively investigated since the pioneering work of Kai Siegbahn.\cite{book_ESCA_molecules} Generally, an increased electron population at the same core ionization center results in a red shift of binding energies. However, less is systematically explored about XAS. Are there similar general trends for absorption energies? In addition to energies, peak intensities are also important. Specifically for vibrationally-resolved XAS, how does the charge state affect the vibronic fine structures?

Recent experimental and theoretical studies on high-resolution X-ray spectra of diatomic systems have covered vibrationally-resolved XAS/XPS spectra for various molecules (N$_2$,\cite{ehara2006symmetry,zhang_franck-condon_2024} CO,\cite{puttner1999vibrationally,zhang_franck-condon_2024} and NO\cite{puttner1999vibrationally,zhang_franck-condon_2024}), cations (NH$^+$,\cite{carniato_vibrationally_2020} N$_2^+$,\cite{lindblad_x-ray_2020,zhang_franck-condon_2024} CO$^+$,\cite{ccouto_carbon_2020,zhang_franck-condon_2024} and NO$^+$\cite{lindblad2022experimental, zhang_franck-condon_2024}), and anion (C$_2^-$\cite{schippers_vibrationally_2023}). To our knowledge, no experimental spectra are currently available for comparison among the three carbon-nitrogen species, necessitating accurate theoretical methods to model the spectra consistently. CN$^+$, CN, and CN$^-$ possess 12, 13, and 14 electrons, respectively, with CN$^+$ and CN$^-$ being closed-shell systems with a singlet ground state (S$_0$), while CN is an open-shell system with a doublet ground state (D$_0$). Multiconfigurational methods are ideal for modeling these systems on the same theoretical footing. In addition to electronic structure methods, accurate vibronic coupling models are essential for yielding precise fine structures. Our recent study \cite{zhang_franck-condon_2024} documented accurate vibrationally-resolved XPS/XAS spectra for various molecules (N$_2$, CO, and NO) and cations (N$_2^+$, CO$^+$, and NO$^+$) at the C/N/O K-edges, aligning well with available high-resolution experimental data. Extensive tests covered different vibronic coupling models, including the harmonic oscillator (HO) approximation and anharmonic potential energy curves (PECs), alongside density functional theory (DFT) and multiconfigurational electronic structure methods. Notably, diatomic systems exhibit greater sensitivity to anharmonic effects, while sensitivity to the electronic structure method is relatively weak \cite{zhang_franck-condon_2024}. This provides a solid foundation for predicting spectra for new systems. 

Predicting spectra is inherently more challenging than referencing systems with existing experimental data. A crucial technical parameter is the accurate determination of absolute core-level transition energies, which poses challenges in quantum chemistry. For multiconfigurational methods, the C/N/O K-edge excitations typically exhibit an accuracy of several eV\cite{hua_mcnox_2024,zhangyu_nonlinear_2016}. The DFT-based $\Delta$Kohn-Sham ($\Delta$KS) method generally achieves an accuracy of 0.5-1.0 eV for 1s excitations of light elements \cite{bagus_consequences_2016,takahashi_functional_2004,du_theoretical_2022}. To obtain complementary information, both multiconfigurational and DFT methods are included in this study. Additionally, we selected CO${}^+$, which has high-resolution experimental data \cite{ccouto_carbon_2020}, as a reference for calibration. For each theoretical method, an ad hoc shift $\delta$ was determined for CO${}^+$, assumed transferable to the carbon-nitrogen species. Notably, CO$^+$ is isoelectronic with CN, making it interesting to study the effects of N$\leftrightarrow$O${}^+$ replacement.

Beyond their implications in molecular physics, CN, CN$^+$, and CN$^-$ are also pivotal in astronomical contexts as interstellar molecules or ions.  The CN ($X^2\Sigma^+$) radical was the second molecule detected in space, after methylidyne (CH) \cite{mckellar_evidence_1940}. In 1995, Bakker et al. \cite{bakker_detection_1995} discovered CN and C$_2$ absorption lines in secondary stars. CN is particularly noteworthy as its millimeter-wave line emission is widely used to probe dense molecular gas and photon-dominated regions of the galactic interstellar medium (ISM) \cite{boger_cn_2005}. Observations of characteristic CN absorption lines in interstellar dust and gas can reveal the abundance and distribution of chemical elements in space. The cyanogen ion, CN$^+$, was discovered in comets through ultraviolet transitions by Snyder et al. in 1992 \cite{snyder_cometary_1992}, and the Goddard High Resolution Spectrograph (GHRS) on the Hubble Space Telescope detected CN$^+$ absorption lines in the interstellar environment \cite{savage_interstellar_1996}. Given the prevalence of CN radicals in space, CN$^+$ is a promising candidate for future radio astronomy searches, playing a significant role in interstellar chemical processes linked to the physical conditions and chemical composition of interstellar dust clouds. The relevance of the corresponding anion, CN$^-$, to astrochemistry has been considered since the 1970s \cite{dalgarno_formation_1973,sarre_possible_1980}, with laboratory observations of rotational spectroscopy achieved in 2000 \cite{gottlieb_rotational_2007,amano_extended_2008}. CN$^-$ is the first diatomic anion detected in astronomy and has garnered significant attention due to its unique presence and mechanisms in interstellar space \cite{agundez_astronomical_2010}. Currently, most molecular anions detected in interstellar and circumstellar gases are heavier linear carbon chains composed of three or more carbon atoms. Thus, studying the abundance of these anions should prioritize shorter anions, particularly CN$^-$, which forms very slowly via radiative electron attachment.

A better understanding of their electronic structure and spectroscopy is vital for X-ray astronomical studies\cite{tielens_molecular_2013}. Additionally, the necessity of studying CN$^+$, CN, and CN$^-$ arises from the fact that the CN group serves as a crucial building block for larger interstellar or circumstellar molecules/ions, such as HCN \cite{snyder1971observations}, HNC \cite{schilke2003first}, CCN \cite{anderson2014detection}, FeCN \cite{zack2011detection}, KCN \cite{ziurys2006chemistry}, MgCN \cite{ziurys2006chemistry}, MgNC \cite{ziurys2006chemistry}, NaCN \cite{ziurys2006chemistry}, SiCN \cite{guelin2000astronomical},   SiNC \cite{guelin2004detection}, H$_2$CN \cite{ohishi1994detection}, H$_2$CN$^+$ \cite{smith1988formation}, CNCN \cite{agundez2018discovery}, CH$_3$CN \cite{remijan2005interstellar}, CH$_3$CN \cite{remijan2005interstellar} and C$_3$H$_7$CN \cite{guelin2000astronomical}. A thorough investigation of these three diatomic species will provide fundamental insights for future studies of more complex interstellar or circumstellar species.

This study aims to simulate high-resolution, vibrationally-resolved XAS spectra of CN, CN$^+$, and CN$^-$ at the carbon K-edge, which will be useful for future identifications in astronomical, physical, and chemical studies. Another theme of this study is to elucidate the influence of charge state, or valence-shell ionization, on the XAS spectrum. In addition to the spectra, we will systematically investigate the electronic structures of these species at both the ground state (GS) and C1s core-excited states, including energy level diagrams, molecular orbital (MO) analysis, and PECs. Understanding their electronic structures is fundamental for interpreting their spectroscopic and various other properties. Our calculations will encompass DFT with the full core hole (FCH) approximation and multiconfigurational methods, including state-averaged restricted active space self-consistent field (SA-RASSCF) \cite{malmqvist_restricted_1990, delcey_analytical_2015, stalring_analytical_2001} and multi-state second-order perturbation theory restricted active space (MS-RASPT2) \cite{finley_multi-state_1998, granovsky_extended_2011}.

\section{Computational details} 
\subsection{Ground-state electronic structure by DFT}

CN$^+$, CN, CN$^-$, and CO$^+$ were oriented along the $Z$ axis, with carbons aligned in the same direction. All geometries were relaxed by DFT with the BLYP functional \cite{lee_development_1988, becke_density-functional_1988} via the GAMESS-US package  \cite{schmidt_general_1993,gordon_chapter_2005}. The IGLO-III basis set \cite{diehl_iglo-method_1990} was set for carbon and aug-cc-pVTZ \cite{kendall_electron_1992,dunning_gaussian_1989} for nitrogen or oxygen,  ensuring consistency with subsequent X-ray spectral calculations with DFT. An energy level diagram was generated at each optimized geometry, visualizing and assigning all occupied and low-lying virtual MOs.

\subsection{(Electronic-only) XAS}

At the optimized GS geometry of each system, the C1s XAS spectrum was simulated using DFT, RASSCF, and RASPT2. DFT calculations employed the full core hole approximation \cite{triguero_calculations_1998}.  For multiconfigurational calculations, state-averaged (SA) CASSCF  over 5 valence states was first performed, with the active space denoted as CAS($N_\text{el}$, $o$), where $N_\text{el}$ and $o$ represent the number of active electrons and orbitals, respectively. Specifically, 12 active orbitals were used, with $N_\text{el}$ set to 10/11/12 for CN$^+$/CN,CO$^+$/CN$^-$. While only the ground state is directly related to XAS, it is standard practice to include additional valence-excited states for state averaging. This was followed by SA-RASSCF calculations over 30 C1s excited states ($n_\text{c}$=30). The usage of a few tens, and sometimes a few hundreds, of core-excited states is to cover an energy range of several to a few tens of eVs as used in other systems and various edges .\cite{zhangyu_nonlinear_2016, hua_study_2016, hua_jpcl_2019, pinjari_cost_2016, temperton_site-selective_2020, gdelcey_soft_2022} Consistent active space was selected for each system: CN$^+$, RAS(10, 1/11/0); CN and CO$^+$, RAS(11, 1/11/0); and CN$^-$, RAS(12, 1/11/0). The notation RAS($N_\text{el}$, $o_1$/$o_2$/$o_3$) followed previous studies \cite{hua_monitoring_2016, zhangyu_nonlinear_2016, hua_study_2016, hua_jpcl_2019},   where $o_1$/$o_2$/$o_3$ denotes sizes of the RAS1/RAS2/RAS3 spaces, respectively, with RAS1 fixed to the C1s orbital with one electron. These calculations were performed using the Molpro package \cite{werner_molpro_2012}, with the aug-cc-pVTZ basis set \cite{kendall_electron_1992,dunning_gaussian_1989}  for both atoms. Oscillator strengths were then computed from the wavefunctions of the initial and final states using our MCNOX code \cite{mcnox, hua_mcnox_2024}. To validate results and assess dynamic correlation effects, additional multiconfigurational simulations were conducted with the openMOLCAS software \cite{li_manni_openmolcas_2023,fdez_galvan_openmolcas_2019,aquilante_modern_2020}, employing both the SA-RASSCF \cite{malmqvist_restricted_1990, delcey_analytical_2015, stalring_analytical_2001} and MS-RASPT2 \cite{finley_multi-state_1998, granovsky_extended_2011} methods, with the same basis set. Oscillator strengths were computed via the RAS state-interaction (RASSI) approach  \cite{malmqvist1989casscf, malmqvist2002restricted}. For each method, the theoretical spectrum of CO$^+$ was uniformly shifted by $\delta$ to align the first resolved peak to the experiment \cite{ccouto_carbon_2020}. This shift was then applied to calibrate the spectra of all other systems. A half-width at half-maximum (HWHM) value of 0.4 eV was used for spectral broadening across all four systems.

The computation is especially more challenging for the anion than the cation or neutral molecule. In this study, special treatment was applied for CN$^-$ while simulating the electronic-only C1s XAS spectrum.  The resulting aug-cc-pVTZ spectrum, as obtained from the above computational procedure, covers only a narrow region with a single major peak (Fig. \si{S1}). Although we attempted to increase $n_\text{c}$ from 30 to 100, we encountered convergence issues. Therefore, we opted to use a smaller cc-pVDZ basis set \cite{kendall_electron_1992} (keeping all other parameters unchanged) to generate the spectral profile, which was calibrated to the main peak of the aug-cc-pVTZ spectrum (see Fig. \si{S1}). In this sense, we compromised to get both an accurate energy position and acceptable overall shape for the XAS spectrum of CN$^-$, just to compare with the wide energy spectra of CN and CN$^+$. Since our main focus in this work is the vibronic fine structure of the first peak, this does not influence consequent discussions, all of which were conducted using the more accurate aug-cc-pVTZ basis set.

\subsection{Multiconfigurational PECs}

Multiconfigurational potential energy curves for CN$^+$/CN/CN$^-$  were simulated at varying bond distances, ranging from 0.7 to 1.22 {\AA} in increments of 0.04 {\AA}, and from 1.22 to 2.42 {\AA} in increments of 0.1 {\AA}. At each bond distance, valence states were first simulated with all theoretical parameters set to those of the ground state geometry. Results at all structures generate the valence-state PECs. Subsequently, vertical core-excited state PECs were generated, maintaining the same parameters as the ground state geometry, except that only the 5 lowest C1s core-excited states were simulated. Besides SA-RASSCF calculations with Molpro \cite{werner_molpro_2012},  for validation, the openMOLCAS software \cite{li_manni_openmolcas_2023,fdez_galvan_openmolcas_2019,aquilante_modern_2020} was also employed to simulate the PECs with both the SA-RASSCF  \cite{malmqvist_restricted_1990,delcey_analytical_2015,stalring_analytical_2001} and MS-RASPT2 \cite{finley_multi-state_1998,granovsky_extended_2011} methods.  The aug-cc-pVTZ \cite{kendall_electron_1992,dunning_gaussian_1989} was always used.

Multiconfigurational PECs for CN$^+$/CN/CN$^-$ were simulated at varying bond distances, ranging from 0.7 to 1.22 {\AA} in increments of 0.04 {\AA}, and from 1.22 to 2.42 {\AA} in increments of 0.1 {\AA}. At each snapshot, valence states were first simulated with all parameters set the same as the GS geometry. Subsequently, vertical PECs were generated, maintaining the same GS parameters but simulating only the 5 lowest C1s core-excited states. For validation, in addition to SA-RASSCF calculations with Molpro \cite{werner_molpro_2012}, the openMOLCAS software  \cite{li_manni_openmolcas_2023,fdez_galvan_openmolcas_2019,aquilante_modern_2020} was utilized to simulate the PECs using both the SA-RASSCF  \cite{malmqvist_restricted_1990,delcey_analytical_2015,stalring_analytical_2001} and MS-RASPT2  \cite{finley_multi-state_1998,granovsky_extended_2011} methods. The aug-cc-pVTZ basis set  \cite{kendall_electron_1992,dunning_gaussian_1989} was consistently employed throughout. Each PEC is fitted to the Morse potential \cite{morse_diatomic_1929}:
\begin{equation}
\label{eq:morse}
  E(R)=T_{\text{e}}+D_{\text{e}}[1-e^{-\alpha (R-R_{\text{e}})}]^2.  
\end{equation}
Here, $R_{\text{e}}$ denotes the equilibrium internuclear distance, $D_{}\text{e}$ stands for the dissociation energy (well depth),  $\alpha$ = $\sqrt{k_{\text{e}}/2D_{\text{e}}}$ represents a potential width parameter with $k_{\text{e}}$=$[\frac{d^{2}U(R)}{dR^{2}}]R={R_{\text{e}}}$ being the force constant at $R_{\text{e}}$, and $T_{\text{e}}$ is a constant term. Raw data of all PECs are provided in the Supplementary Material.\cite{si_lijingyu1}

\subsection{Vibrationally-resolved XAS}

The theory for computing vibrationally-resolved XAS using both anharmonic and harmonic methods has been detailed elsewhere \cite{zhang_franck-condon_2024}. Vibrationally-resolved C1s XAS spectra of CN$^+$/CN/CN$^-$ were simulated using the wavepacket method based on multiconfigurational PECs with our XspecTime package \cite{XSpecTime,zhang_franck-condon_2024}. The program reads PECs of the initial and final electronic states, performs Morse fitting and computes the spectrum at 201 discrete points across C-N distances from 0.7 to 2.5 \AA.  The wavepacket was propagated for a duration of 6$\times10^6$ a.u. with a time step of 1.0 a.u. A HWHM lifetime of 0.05 eV for the C1s core hole was applied. 

For comparison, we also simulated the vibrationally-resolved C1s XAS spectra based on the BLYP-DFT electronic structure under the harmonic oscillator (HO) approximation. Geometrical optimizations and vibrational frequency calculations were conducted both at the ground and the lowest C1s excited states using  GAMESS-US \cite{schmidt_general_1993,gordon_chapter_2005}. Subsequent Franck-Condon factor (FCFs) were simulated by using the modified \cite{hua_theoretical_2020} DynaVib package \cite{DynaVib}, which was interfaced with GAMESS-US to read the DFT results, including optimized structures, vibrational frequencies, normal modes, and integrals for both electronic states. Lorentzian broadening was adopted with a HWHM of 0.05 eV.

\section{Results}

\subsection{Ground-state electronic structures}

\begin{figure*}[htbp]
\centering
\includegraphics[width=1\textwidth]{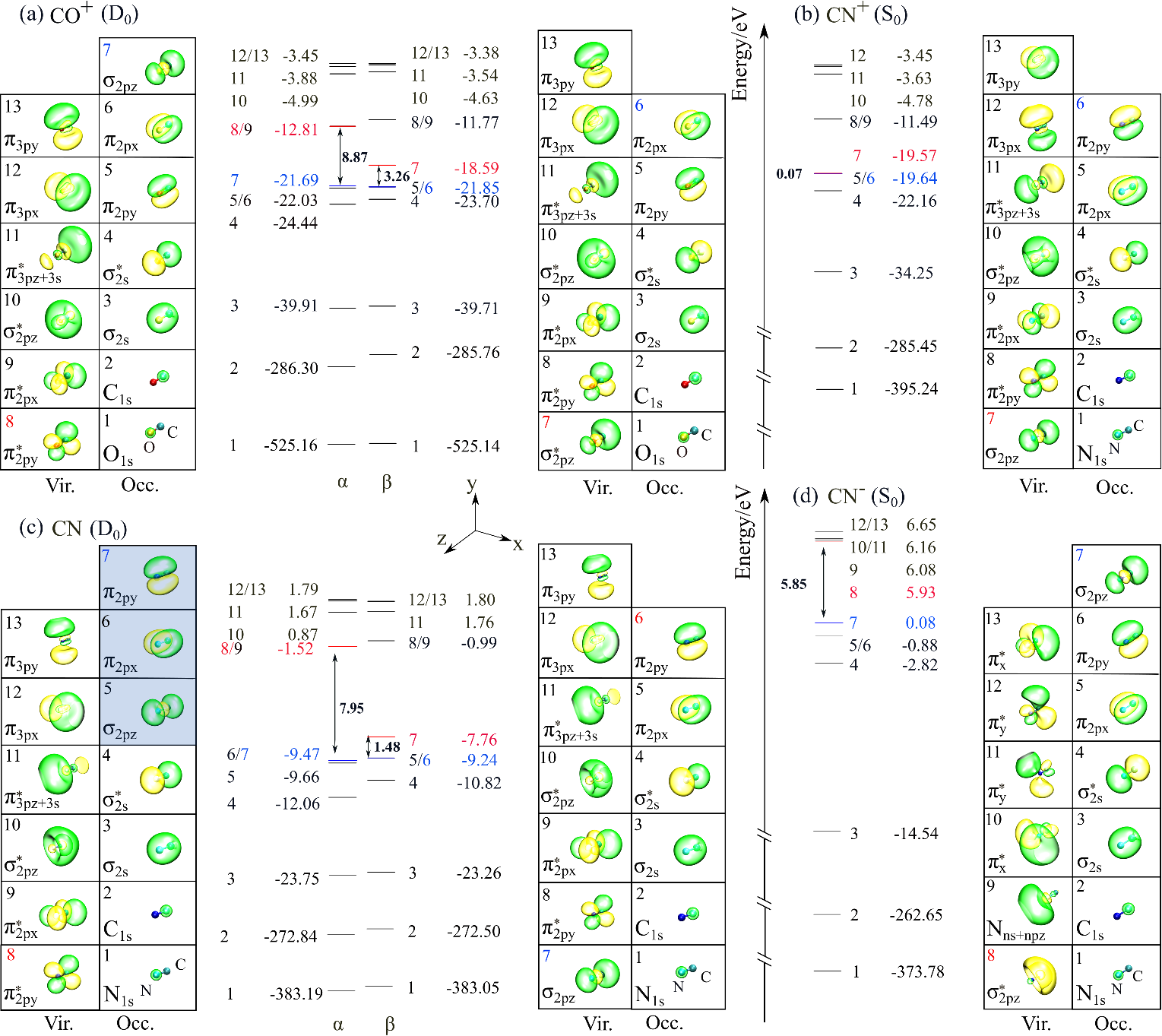}
\caption{Energy level diagram of (a) CO$^+$, (b) CN$^+$, (c) CN, and (d) CN$^-$ in their ground electronic states computed by DFT  with the BLYP functional. In panels (a,c), spin-unrestricted DFT was used for CO$^+$ and CN. Occupied and low-lying virtual MOs are shown in separate columns and interpreted by their major components. Shaded area in CN $\alpha$ occupied orbitals is to show a different order in MOs 5-7 ($\sigma_{2pz}$, $\pi_{2px}$/$\pi_{2py}$) as compared to others ($\pi_{2px}$/$\pi_{2py}$, $\sigma_{2pz}$). Degenerate orbitals are indicated by ``/''. HOMO (blue) -- LUMO (red) gap of each spin is indicated by arrows.  All energies are in eV.
}
\label{fig:Orbital}
\end{figure*}

Figure \ref{fig:Orbital} illustrates the simulated BLYP energy level diagrams for CO$^+$, CN$^+$, CN, and CN$^-$ in their optimized ground-state geometries. Each system exhibits distinct HOMO-LUMO gaps, where HOMO and LUMO refer to the highest occupied and the lowest unoccupied molecular orbitals, respectively. For the two open-shell systems, CO$^+$ has $\alpha$ and $\beta$ gaps of 8.9 and 3.3 eV, respectively [Fig. \ref{fig:Orbital}(a)]. In contrast, its isoelectronic counterpart, CN, shows $\alpha$ and $\beta$ gaps of 8.0 and 1.5 eV, respectively [Fig. \ref{fig:Orbital}(c)]. For the two closed-shell systems, CN$^+$ exhibits an exceptionally small gap of less than 0.1 eV [Fig. \ref{fig:Orbital}(b)], whereas CN$^-$ displays a gap of 5.9 eV [Fig. \ref{fig:Orbital}(d)].

Despite these energy differences, orbital features across various species (and different spins of the same species) exhibit remarkable similarities. This holds for the occupied MOs of CN$^-$ and both the occupied and unoccupied MOs of the other three systems. Specifically, in Fig. \ref{fig:Orbital}(a), MOs 1-10 for both $\alpha$ and $\beta$ spins of CO$^+$ are characterized as follows: O1s, C1s, $\sigma_{2s}$, $\sigma_{2s}^*$, $\pi_{2py}$/$\pi_{2px}$, $\sigma_{2pz}$, $\pi_{2py}^*$/$\pi_{2px}^*$, and $\sigma_{2pz}^*$. (Note that delocalized $\sigma_{1s}$ and $\sigma_{1s}^*$ are absent here as we localized the 1s orbitals, and ``/'' describes degenerate orbitals.) The $\alpha$ and $\beta$ orbitals of the same index are nearly identical. Figures \ref{fig:Orbital} (b,c) show that orbitals of CN$^+$  and $\beta$  orbitals of CN  have similar features and orders to those of CO$^+$. In contrast, the $\alpha$ orbitals of CN present a different sequence for MOs 5-7 ($\sigma_{2pz}$, $\pi_{2py}$/$\pi_{2px}$), differing from the order observed in the orbitals of CO$^+$, CN$^+$,  and the $\beta$ orbitals of CN ($\pi_{2py}$/$\pi_{2px}$, $\sigma_{2pz}$). Figure \ref{fig:Orbital}(d) illustrates that the occupied orbitals (MOs 1-7) of CN$^-$ closely resemble those of CO$^+$, CN$^+$, and the $\beta$ orbitals of CN. However, the unoccupied orbitals are more diffuse and mixed, showing significant differences from others and complicating their major MO feature interpretations. In summary, CN$^-$ has notably different unoccupied orbitals compared to the other diatomic systems due to the presence of an extra electron.

\subsection{XAS spectra}\label{sec:xasresult:ElecOnly}

\begin{figure*}[htbp]
\centering
\includegraphics[width=1\textwidth]{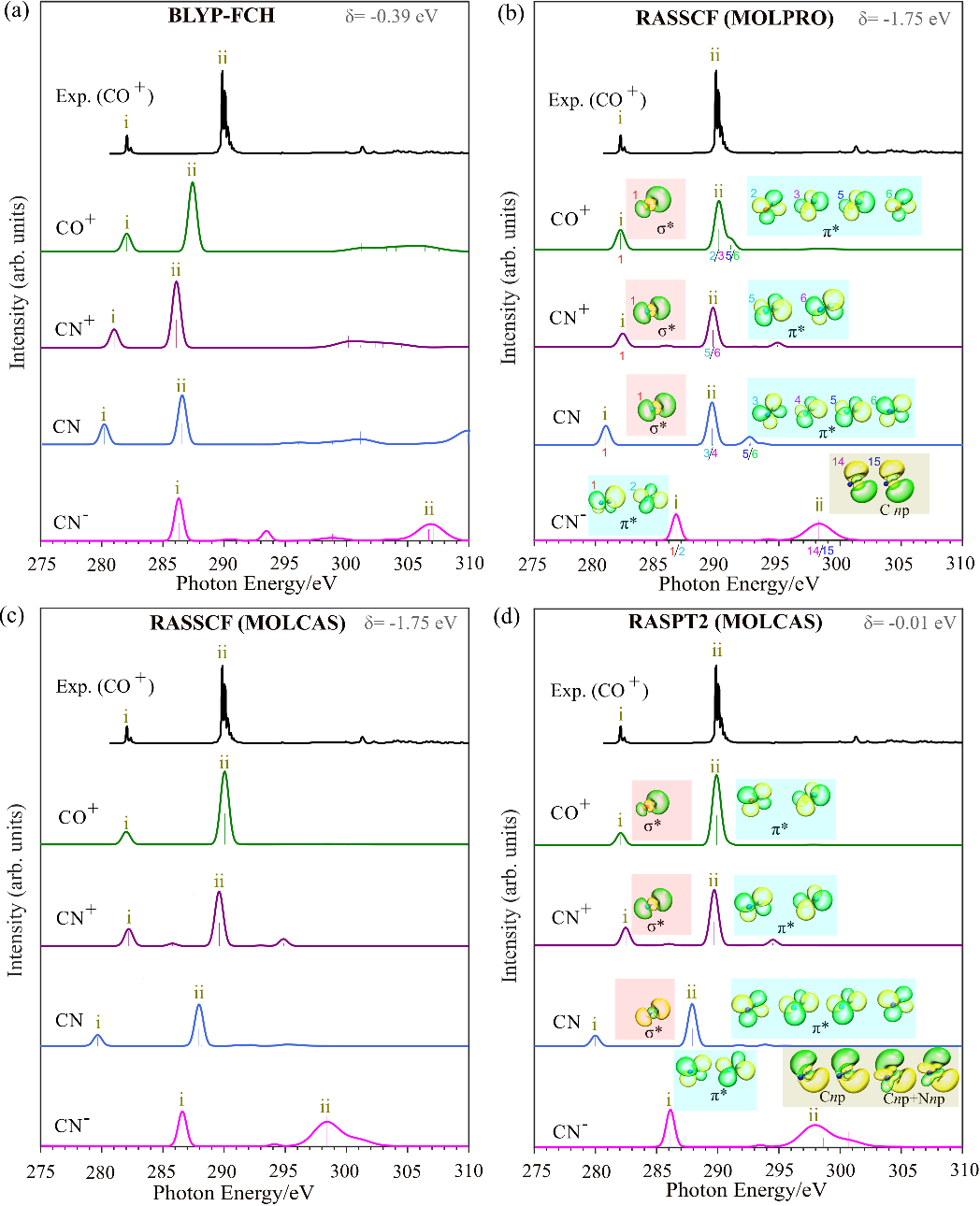}
\caption{Simulated C1s XAS spectra for CO, CO$^+$, CN, CN$^-$, and CN$^+$ at their ground-state geometries, using different theoretical methods and software as indicated. For each method, the theoretical spectrum for CO$^+$ was shifted by $\delta$ to align with experimental data \cite{ccouto_carbon_2020}, which was then applied to all other systems. In panels (b) and (d), major peaks are interpreted by NTOs alongside (note for degeneracy), with their major characteristics assigned and distinguished by three differently colored backgrounds.}
\label{fig:xas:eleconly}
\end{figure*}

Figure \ref{fig:xas:eleconly} presents  C1s XAS spectra of the four diatomic systems at their DFT-optimized ground-state geometries, simulated using DFT, RASSCF, and RASPT2 methods. Detailed positions and assignments for major peaks, labeled as $i$ and $ii$, are provided in Table \ref{tab:peak}. When comparing the absorption energy of the first peak $i$ across the three species, there is no monotonic relationship with charge state. For example, the BLYP method yields absorption energies of 281.0 eV, 280.5 eV, and 286.1 eV for CN$^+$, CN, and CN$^-$, respectively. Similarly, RASPT2 provides values of 282.5 eV, 280.0 eV, and 286.1 eV for these species. One possible explanation for this observation is that peak $i$ arises from different absorption features, as will be discussed further below. This suggests that the relationship between energy and charge state in XAS is more complex than in XPS.

For CO$^+$, the BLYP calculation gives acceptable agreement with the experiment \cite{ccouto_carbon_2020}, reproducing the overall shape, although it underestimates the separation between peaks $i$ and $ii$ (BLYP, 287.5 eV; experiment, 289.9 eV). Three multiconfigurational calculations yield consistent results, showing strong agreement with the experiment and better performance than BLYP regarding the separation between peaks $i$ and $ii$. This comparison with the experiment for CO$^+$ validates the accuracy of our predictions for the three C-N systems.

The spectra of CN$^+$ and CN show similar profiles to CO$^+$, except with redshifts of about 1.2 and 1.8 eV, respectively. In contrast, the CN$^-$ spectrum is distinctly different from the other three, reflecting unique features in the unoccupied region, consistent with the ground electronic state.

Major peaks were interpreted using natural transition orbitals (NTOs) \cite{martin2003natural, malmqvist2012binatural}, focusing on RASSCF [Fig. \ref{fig:xas:eleconly} (b)] and RASPT2 [Fig. \ref{fig:xas:eleconly} (d)] methods as examples. The weak peak $i$ and strong peak $ii$ for CO$^+$, CN$^+$, and CN corresponds to C1s$\rightarrow$$\sigma^*$ and C1s$\rightarrow$$\pi^*$ transitions, respectively. This aligns with the unoccupied-level electronic structure in the ground state, where the $\sigma_{2pz}$ orbital (MO 7; only the $\beta$ spin for CO$^+$ and CN) has lower energy than $\pi^*_{2px}$ and $\pi^*_{2py}$ (MOs 8-9), indicating relatively weak electronic relaxation due to the introduced core hole.

The CN$^-$ spectrum, however, has different assignments. The lowest peak $i$, predicted at 286.11--286.69 eV, corresponds to C1s$\rightarrow\pi^*$ transitions, including transitions to two degenerate $\pi^*$ orbitals (in $xz$ and $yz$ directions). This degeneracy is the reason the first peak in CN$^-$ is roughly twice the intensity of those in CN or CN$^+$, where the C1s$\rightarrow\sigma^*$ transition along the bond length direction lacks degeneracy. In the higher energy region (ca. 298.10--298.82 eV), a broad peak arises from multiple C1s$\rightarrow$C$n$p transitions, each with mixed components (see also \si{Fig. S2} for a more localized view with a larger isovalue). The anion demonstrates the largest core hole effect due to significant differences in the electronic structures of the ground and core-excited states. DFT also reveals substantial differences compared to the three multiconfigurational simulations, highlighting the necessity of explicit static correlation.  Consistency among the three multiconfigurational spectra indicates a weak influence of dynamical correlations.

\subsection{Core-hole induced changes in bond lengths and vibrational frequencies}

Table \ref{tab:frequencies} presents simulated bond lengths for each system in both the optimized ground state ($l'$) and the lowest C1s core-excited state ($l$), along with their difference ($\Delta l \equiv l - l'$), as predicted by BLYP-DFT. It also includes the vibrational frequencies for the initial ($\omega'$) and final ($\omega$) states, with the corresponding change ($\Delta \omega \equiv \omega - \omega'$). The changes in C-N bond lengths induced by C1s excitation for CN$^+$, CN, and CN$^-$ are -5.6, -6.8, and -6.1 pm, respectively, indicating similar magnitudes of shortening across all three species. It is noted that this is quite system-dependent, as in CO$^+$ and CO, the changes induced by the C1s core hole are -3.7 and +2.5 pm, respectively \cite{zhang_franck-condon_2024}.

The ground-state vibrational frequencies range narrowly from 2015.1 to 2070.9 cm$^{-1}$ (55.8 cm$^{-1}$ span), whereas excited-state frequencies increase and vary more widely from 2343.5 to 2467.6 cm$^{-1}$ (124.1 cm$^{-1}$ span), underscoring the strong influence of the core hole. The vibrational frequency changes show mild increases: CN$^+$ at 328.4 cm$^{-1}$, CN at 377.8 cm$^{-1}$, and CN$^-$ at 425.9 cm$^{-1}$. The similar magnitudes parallel with the observed bond length shortening.  Similar absolute shifts of several hundred wavenumbers due to C/N/O 1s core holes have been reported in other diatomic systems \cite{zhang_franck-condon_2024}.

\subsection{Potential energy curves}

\begin{figure*}
\centering
\includegraphics[width=0.95\textwidth]{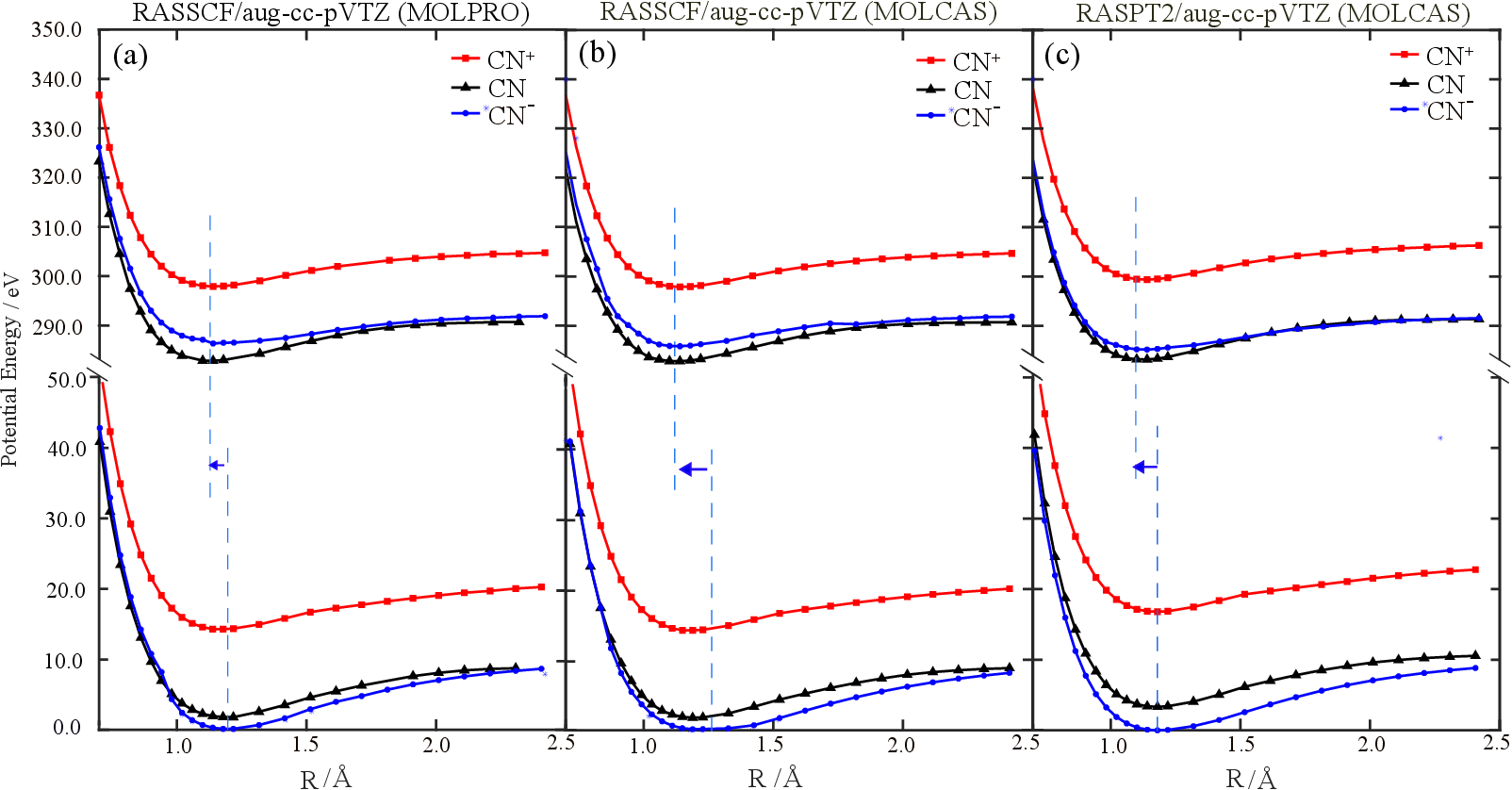}
\caption{Simulated PECs for the ground  (bottom) and the lowest C1s core-excited (top) states of CN$^+$, CN,  and CN$^-$ using different methods (software used is indicated in parentheses) as indicated. In panel (d),  (blue) dashed lines and arrow are to highlight the decrease in equilibrium positions, taking CN$^-$ as an example.
}
\label{fig:pec}
\end{figure*}

Figure \ref{fig:pec} illustrates PECs for the ground and the lowest C1s excited states of CN$^+$, CN, and CN$^-$, simulated using various electronic structure methods. All PECs are well-fitted to Morse potentials, with the resulting parameters detailed in Table \ref{tab:potential}. Within each method, the ground state curves decrease in the order of CN$^+$, CN, and CN$^-$, with CN$^+$ distinctly higher than the other two. In the core-excited state, CN$^+$ retains the highest energy, but the order of the CN and CN$^-$ curves is reversed. This shift arises from the core hole effect, which is significant due to the small energy separation between CN and CN$^+$ in the ground state, altering their relative energies.

The three multiconfigurational methods yield similar PECs, validating our predictions.  In either the ground or excited state, the predicted equilibrium bond lengths for the three species and by all methods show consistent values. For example, RASPT2 simulations predicted ground-state equilibrium bond lengths for CN$^+$, CN, and CN$^-$ of 1.188, 1.176, 1.180 {\AA}, respectively (Table \ref{tab:potential}). These values also align well with the optimized bond lengths of 1.183 {\AA} (CN$^+$), 1.183 {\AA} (CN), and 1.172 {\AA} (CN$^-$) obtained using harmonic oscillator approximations and BLYP-DFT (Table \ref{tab:frequencies}).

Bond length shortening is illustrated both in Fig. \ref{fig:pec} (highlighting CN$^-$) and Table \ref{tab:potential}. The reduction of 4--6 pm due to the core hole effect is consistent with the 6--7 pm observed with harmonic oscillator approximations (Table \ref{tab:frequencies}).

In the ground state, the fitted dissociation energies ($D_e$) increase across the series: CN$^+$ (6.3--6.4 eV), CN (8.1--8.2 eV), and CN$^-$ (9.4--10.3 eV) as shown in Table \ref{tab:potential}. While in the excited state, dissociation energies change significantly: CN$^+$ (0.27 eV), CN (0.33 eV), and CN$^-$ (0.22--0.25 eV), with CN now exhibiting the largest dissociation energy. These results highlight the strong influence of the core hole effect.


\subsection{Vibronic fine structures}

\begin{figure*}
\centering
\includegraphics[width=0.95\textwidth]{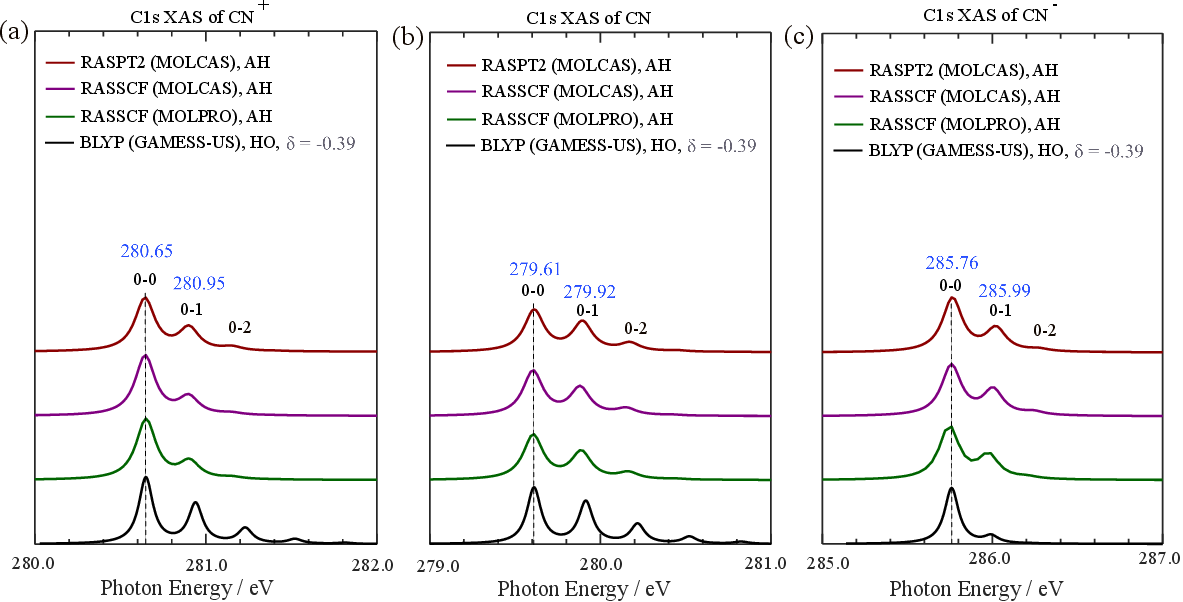}
\caption{Vibrationally-resolved C1s XAS spectra for the lowest C1s excited state of (a) CN$^+$, (b) CN, and (c) CN$^-$ simulated using various theoretical methods: the HO method with the BLYP-DFT electronic structure (black) and the AH method with the multiconfigurational PECs (colored). To facilitate comparison with future experiments, an \textit{ad hoc} shift of $\delta = 0.39$ eV [derived from Fig. \ref{fig:xas:eleconly}(a)] was applied to the BLYP spectrum. For clarity in fine structure comparisons, all other methods were aligned to the BLYP spectrum by adjusting the 0-0 peak. Curves of each system is respectively normalized.} \label{fig:xas:vibronic}
\end{figure*}

Figure \ref{fig:xas:vibronic} presents simulated vibronic fine structures for the lowest C1s excited states of CN$^+$, CN,  and CN$^-$, derived from DFT and multiconfigurational methods. For each system, the three multiconfigurational methods yield nearly equivalent spectral profiles, validating the computational accuracy and indicating a weak influence of dynamic electron correlation on vibronic coupling. The most significant difference in the predicted spectra of the three systems lies in the energy positions of the 0-0 peaks: 280.7, 279.6, and 285.8 eV for CN$^+$, CN, and CN$^-$, respectively. While the profiles appear generally similar, each is dominated by the 0--0 peak, followed by a weaker 0--1 peak.  The intensity ratio of the 0--1 peak to the 0--0 peak, denoted as $F_{01}/F_{00}$ \cite{hua_theoretical_2020}, is approximately 0.55, 0.75, and  0.49, respectively.  Additionally, the predicted energy difference between the 0--0 and 0--1 peaks is  $\sim$0.23 eV for CN$^-$ and $\sim$0.30 eV for CN$^+$ and CN. In the region beyond, an even weaker 0--2 peak is still identifiable in CN, but less visible in CN$^-$ or CN$^+$. The similarity in profiles results from the resemblance of PECs (Fig. \ref{fig:pec}). The dominance of the 0--0 peak suggests the relatively small displacement of the PECs, consistent with bond length changes of 4-6 pm (Fig. \ref{fig:pec} and Table \ref{tab:potential}).

The DFT curves exhibit a rough resemblance to the corresponding multiconfigurational results, yet they contain significant discrepancies. The reduced accuracy of the DFT results primarily arises from the harmonic oscillator approximation used. A previous study \cite{zhang_franck-condon_2024} highlighted the pronounced sensitivity of anharmonic effects and the diminished responsiveness of electronic structure methods across several diatomic systems.

\begin{figure*}
\centering
\includegraphics[width=0.5\textwidth]{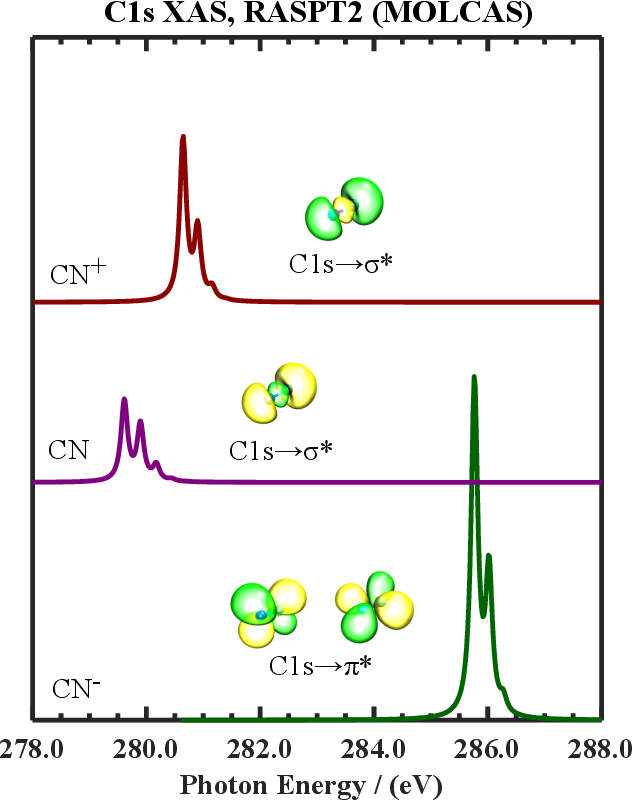}
\caption{Overlay of vibrationally-resolved C1s XAS spectra for the lowest C1s excited state of  CN$^+$, CN, and CN$^-$, simulated using the anharmonic method with RASPT2 PECs (see top curves in Fig. \ref{fig:xas:vibronic}). Note that the same scale for the Y-axis is used across the three systems. Each normalized curve in Fig. \ref{fig:xas:vibronic} has been  multiplied by the corresponding electronic oscillator strengths ($f$) [see leftmost sticks in Fig. \ref{fig:xas:eleconly}(d)] and the degeneracy. Insets recaptures the electronic transition, indicating degeneracies of 1, 1, and 2, respectively.} \label{fig:xas:vibronic:overlay}
\end{figure*}

In Fig. \ref{fig:xas:vibronic:overlay}, we superimpose the RASPT2 vibronic profiles for the three systems to facilitate a direct comparison. Each normalized vibronic profile, shown in Fig. \ref{fig:xas:vibronic}, has now been multiplied by the computed electronic oscillator strength ($f$), as indicated by the sticks in Fig. \ref{fig:xas:eleconly}. The significantly higher intensity in the CN$^-$ curve results from the two-fold degeneracy of the $\pi^*$ orbitals  for the C1s$\rightarrow\pi^*$ transitions, as discussed in Section \ref{sec:xasresult:ElecOnly}. 

\section{Discussions on valence shell ionization potentials}

The three species are interconnected through two ionization processes (\ref{eq:i1})-(\ref{eq:i2}). Using the equilibrium distances from multiconfigurational PECs, we also computed the valence-shell vertical ($V$) and adiabatic ($A$) ionization potentials for CN$^-$ and CN. Specifically, the vertical and adiabatic IPs for the ionization process (\ref{eq:i2}), for instance, are defined as the potential energy differences between CN ($E_{\text{CN}}$) and CN$^+$ ($E_{\text{CN}^+}$): 
\begin{eqnarray}
V &=& E_{\text{CN}^+}(R_{\text{e,CN}}) - E_{\text{CN}}(R_{\text{e,CN}}) \\
A &=& E_{\text{CN}^+}(R_{\text{e,CN}^+}) - E_{\text{CN}}(R_{\text{e,CN}}).
\end{eqnarray}
Here, $R_{\text{e,CN}}$ and $R_{\text{e,CN}^+}$ denote the equilibrium distances.

Table \ref{tab:IP} presents the simulated  IPs results. In the ground state, the close agreement of the adiabatic and vertical IP values indicates minimal structural rearrangement upon ionization. The RASPT2-predicted IPs for processes (\ref{eq:i1}) and (\ref{eq:i2}) are 3.6 eV and 13.7 eV, respectively, illustrating that ionizing a second valence electron from CN$^-$ is significantly more challenging than the first. This aligns with chemical intuition: the extra electron in the anion is more loosely bound, while the outer electron in the neutral molecule experiences a larger effective nuclear charge, resulting in tighter binding. The RASSCF results are consistent but slightly underestimate the IPs, yielding values of 1.7–2.2 eV and 12.6–12.7 eV, reflecting the influence of dynamic correlation. 

Additionally, Table \ref{tab:IP} includes simulated IPs for the lowest C1s excited state, describing the shake-off process observed in XAS, where primary core excitation leads to the ionization of outer electrons. At the RASPT2 level, the computed vertical (adiabatic) IPs are -1.6 (-1.7) eV and 16.2 (16.1) eV, respectively. Typically, IPs are positive; the negative IP for CN$^-$ suggests a lower likelihood of the shake-off process, indicating a less stable final state, while the positive IP for CN indicates a higher probability for this process. Similar to the ground state, RASSCF yields values comparable to RASPT2.

\section{Summary and conclusions}

In summary, we have theoretically predicted accurate C1s XAS spectra for CN$^+$, CN, and CN$^-$ through multiconfigurational and DFT simulations, providing valuable references for future X-ray spectroscopic studies in astronomy and chemical physics. Due to the lack of experimental data, CO$^+$, an isoelectronic species of CN, was used for calibration and validation with various electronic structure methods. Our analysis reveals that there is no monotonic relationship between the absorption energy of the first peak $i$ and charge state when comparing the three species. This complexity suggests that the relationship between energy and charge state in XAS is more intricate than in XPS. Notably, the first electronic peak for CO$^+$, CN$^+$, and CN is assigned to C1s$\rightarrow\sigma^*$ transitions, while that for CN$^-$ corresponds to a C1s$\rightarrow\pi^*$ transition. We further explained that the two-fold degeneracy ($\pi^*_{xz}$ and $\pi^*_{yz}$, with the system oriented along the $z$ axis) of C1s$\rightarrow\pi^*$ transitions  is the reason why CN$^-$ exhibits a significantly stronger first absorption compared to the other two systems.

Vibronic fine structures for the lowest C1s transition, well-separated from higher absorptions, were computed using RASSCF- and RASPT2-level anharmonic PECs via the quantum wavepacket method, yielding consistent results. The 0-0 absorption energies are distinctly different at 280.7, 279.6, and 285.8 eV for the three systems. Although the vibronic profiles are similar, they differ in the intensity ratios of the 0-0 and 0-1 peaks. Spectra were assigned and correlated with ground-state electronic structures to investigate the core hole effect. PECs indicate a decrease of 4-6 pm in equilibrium bond lengths due to the C1s core hole for all species. Additionally, using the harmonic approximation with DFT, the C1s core hole leads to a bond length reduction of 6-7 pm and an increase in vibrational frequency of 300–400 cm$^{-1}$. The resulting spectra closely match the vibronic fine structures from anharmonic calculations, underscoring the influence of anharmonicity. These findings enhance our understanding of the electronic structure, X-ray spectra, and vibronic coupling effects, illustrating the impact of charge state. Our group is also actively working on high-resolution X-ray emission and photoelectron spectra of molecules, aiming to establish a comprehensive theoretical library that may support molecular physics, chemical studies, and X-ray astronomical research.

\section*{Acknowledgments}
Financial support from the National Natural Science Foundation of China (Grant No. 12274229) is greatly acknowledged. Thanks to Prof. Jicai Liu for providing the Molpro software for part of the calculations.

%

\clearpage
\begin{table}[!htb]
    \caption{
Energies and assignments for major peaks (defined in Fig. \ref{fig:xas:eleconly}) in the C1s XAS spectra of CO$^+$, CO, CN$^+$, CN, and CN$^-$, simulated using different methods and software at their ground-state equilibrium geometry. Peak energies here have been calibrated by adding a method-specific \emph{ad hoc} shift $\delta$ (see values in Fig. \ref{fig:xas:eleconly}), obtained by aligning peak $i$ of CO$^+$ to the corresponding experiment \cite{ccouto_carbon_2020}. All energies are in eV, and $\Delta$ denotes difference between peaks $ii$ and $i$.
} 
\label{tab:peak}
\hspace{0.1cm}
\begin{tabular}{@{\hskip 2pt}l @{\hskip 2pt}l @{\hskip 2pt}l @{\hskip 2pt}l @{\hskip 2pt}l}
        \hline\hline
            Cmpd. & Method & Peak $i$ & Peak $ii$ & $\Delta$\tnote{a}   \\ \hline
            CO$^+$ & BLYP-FCH & 282.06 & 287.46 &  5.40  \\ 
            & RASSCF (MOLPRO) & 282.06 &  289.90 & 7.84   \\ 
            & RASSCF (MOLCAS) & 283.06 &  290.11 & 7.05   \\ 
            & RASPT2 (MOLCAS) & 282.06 &  289.88 &  7.82 \\ 
            & Expt. & 282.06  &  289.88 &   7.82 \\ 
            & Assignment & C1s$\rightarrow\sigma^*$ & C1s$\rightarrow\pi^*$ &   \\ \hline
            CN$^+$ & BLYP-FCH & 281.01 & 286.11 &   5.10 \\ 
            & RASSCF (MOLPRO) & 282.22 &  289.50 &  7.28   \\ 
            & RASSCF (MOLCAS) & 282.24 &  289.61 &  7.37   \\ 
            & RASPT2 (MOLCAS) & 282.46 &  289.72 &  7.26   \\ 
            & Assignment & C1s$\rightarrow\sigma^*$ & C1s$\rightarrow\pi^*$ &   \\ \hline
            CN & BLYP & 280.50 & 286.85 &  6.35   \\ 
            & RASSCF (MOLPRO) & 279.76 &  288.78 &  9.02  \\ 
            & RASSCF (MOLCAS) & 279.71 &  287.98 &  8.27   \\ 
            & RASPT2 (MOLCAS) & 279.99 &  287.92 &   7.93  \\ 
            & Assignment & C1s$\rightarrow\sigma^*$ & C1s$\rightarrow\pi^*$ &  \\ \hline
            CN$^-$ & BLYP-FCH & 286.11 & 298.65 &  12.54  \\ 
            & RASSCF (MOLPRO) & 286.69 &  298.36 & 11.67   \\ 
            & RASSCF (MOLCAS) & 286.68 &  298.82 &  12.14   \\ 
            & RASPT2 (MOLCAS) & 286.12 &  298.10 &  11.98   \\ 
            & Assignment & C1s$\rightarrow\pi^*$ & C1s$\rightarrow$$p$Ryd &   \\
        \hline
\end{tabular}
\end{table}

\begin{table}[htb!]
  \begin{center}
    \caption{
Comparison of simulated vibrational frequencies $\omega$ (in cm$^-{}^1$) and optimized bond lengths  (in \AA) for ground state ($\omega'$, $l'$) and the lowest C1s excited state ($\omega$, $l$) for CN$^+$, CN, and CN$^-$ calculated by DFT with the BLYP functional. $\Delta \omega=\omega-\omega'$ and $\Delta l=l-l'$ represent the frequency and bond length changes, respectively.
} \label{tab:frequencies}
    \begin{tabular}{lllcccccc}
    \hline\hline 
        Cmpd.&$\omega'$&$\omega$& $\Delta\omega$&$l'$&$l$&$\Delta l$\\ \hline
                CN$^+$ & 2015.1& 2343.5& +328.4 & 1.183 & 1.127 & -0.056 \\ 
        CN & 2070.9& 2448.7& +377.8 & 1.182 & 1.114 & -0.068  \\ 
        CN$^-$ & 2041.7 &  2467.6 & +425.9 & 1.172 & 1.111 & -0.061\\ 
        \hline\hline 
    \end{tabular} 
\end{center}
\end{table}

\begin{table*}[ht]
    \centering
    \caption{Fitted Morse \cite{morse_diatomic_1929} parameters [see $T_{\text{e}}$, $D_{\text{e}}$, $\alpha$, and $R_{\text{e}}$ in Eq (\ref{eq:morse})] for the potential energy curves of CN$^+$, CN, and CN$^-$, obtained using various electronic structure methods (software used is indicated in parentheses). Superscripts 0 and 1 label the ground and the lowest C1s excited state, respectively.}
\label{tab:potential}
    \begin{tabular}{lcllccccccccc}
    \hline\hline
        \multirow{2}{*}{Cmpd.}&\multirow{2}{*}{Method} &\multicolumn{4}{c}{Ground state} &\multicolumn{4}{c}{Lowest C1s excited state} &\multirow{2}{*}{$\Delta R_{\text{e}}$ (\AA)} \\ 
        \cmidrule(l){3-6}\cmidrule(l){7-10} 
        & &  $T_{\text{e}}^0$ (eV)& $D_{\text{e}}^0$ (eV) & $\alpha^0$  (a.u.)& $R_{\text{e}}^0$ (\AA) & $T_{\text{e}}^1$ (eV)& $D_{\text{e}}^1$ (eV) & $\alpha^1$ (a.u.)& $R_{\text{e}}^1$ (\AA) &\\ 
        \midrule
         CN$^+$ &RASSCF (MOLPRO)&0.054&6.343&1.359&1.184 &283.543&0.266&1.419&1.146&-0.038\\
        &RASSCF (MOLCAS)&0.054&6.367&1.353&1.184 &283.815&0.266&1.423&1.146&-0.038\\
        &RASPT2 (MOLCAS)&0.082&6.367&1.345&1.188 &282.454&0.268&1.434&1.141&-0.047\\ \hline 
        CN&RASSCF  (MOLPRO)&0.136&8.191&1.280&1.181&281.366&0.332&1.409&1.128 &-0.053\\
        &RASSCF (MOLCAS)&0.054&8.055&1.280&1.183 &281.366&0.326&1.394&1.129&-0.054\\
        &RASPT2 (MOLCAS)&0.082&8.218&1.288& 1.176&280.005&0.332&1.418&1.146&-0.056\\ \hline 
       CN$^-$ &RASSCF (MOLPRO)&0.082&10.286&1.140&1.218 &286.264&0.216&1.441&1.171&-0.047\\
        &RASSCF (MOLCAS)&-0.435&9.415&1.195&1.203 &285.992&0.237&1.456&1.154&-0.049\\
        &RASPT2 (MOLCAS)&-0.136&10.123&1.209&1.180 &284.903&0.250&1.481&1.136&-0.044\\
        \hline
    \end{tabular}
\end{table*}

\begin{table}[ht]
    \centering
    \caption{
Adiabatic ($A^0$ and $A^1$) and vertical ($V^0$ and $V^1$) ionization potentials (in eV) for two valence electron ionization processes: (1) CN$^-$$\rightarrow$CN and (2) CN$\rightarrow$CN$^+$, calculated using various electronic structure methods (software indicated in parentheses). Superscripts 0 and 1 denote the ground and the lowest C1s excited states, respectively. Negative values (highlighted in parentheses) indicate non-physical shake-off processes; see text for details.
} 
\label{tab:IP}
\begin{tabular}{@{\hskip 2pt}l @{\hskip 2pt}l @{\hskip 2pt}c @{\hskip 2pt}c @{\hskip 2pt}c @{\hskip 2pt}c}
    \hline\hline
    \multirow{2}{*}{No.} & \multirow{2}{*}{Method} & \multicolumn{2}{c}{Ground state} & \multicolumn{2}{c}{C1s excited state} \\\cmidrule(l){3-4}\cmidrule(l){5-6} 
    & & $V^0$ & $A^0$ & $V^1$ & $A^1$ \\ \midrule
    1 & RASSCF (MOLPRO) & 1.807 & 1.734 & (-3.286) & (-3.378) \\
      & RASSCF (MOLCAS) & 2.186 & 2.166 & (-2.966) & (-2.990) \\
      & RASPT2 (MOLCAS) & 3.629 & 3.629 & (-1.646) & (-1.655) \\ \hline 
    2 & RASSCF (MOLPRO) & 12.600 & 12.598 & 15.086 & 15.076 \\
      & RASSCF (MOLCAS) & 12.653 & 12.653 & 15.034 & 15.025 \\
      & RASPT2 (MOLCAS) & 13.719 & 13.715 & 16.164 & 16.146 \\
    \hline
\end{tabular}
\end{table}


\end{document}